\documentclass[final]{IEEEtran}

\usepackage{graphicx}  
\usepackage{amsthm}
\usepackage{amssymb}
\usepackage{tikz}
\usetikzlibrary{positioning,calc,patterns}
\usepackage{pgfplots}
\usepgfplotslibrary{fillbetween}
\usepackage{amsmath}
\usepackage{comment}
\usepackage{url}
\usepackage{hyperref}
\usepackage{todonotes}
\usepackage{subcaption}

\newcommand{\diag}{\operatorname{diag}}

\newcommand{\be}{\begin{equation}}
\newcommand{\ee}{\end{equation}}
\newcommand{\R}{\mathbb R}

\newcommand{\N}{\mathbb N}

\newtheorem{Theorem}{Theorem}
\newtheorem{Proposition}{Proposition}

\newtheorem{Corollary}{Corollary}

\newtheorem{Remark} {Remark}
\newtheorem{Example} {Example}
\newcommand{\such}{\, | \,}

\title{A networked  small-gain theorem based on discrete-time diagonal stability 
\thanks{The research of RO is partly supported by a  Khazanov Scholarship and the Air Force Office of Scientific Research under Grant No.~FA9550-23-1-017. The research of MM is partially supported by a research grant from  the Israeli Science Foundation~(ISF).  }}

\author{Ron Ofir and Michael Margaliot
\thanks{RO was with the School of Elec. Eng., Tel-Aviv University, Tel Aviv~69978, Israel, and is now with the Department of Electrical \& Computer Engineering, Yale University, New Haven CT~06520, USA. MM (Corresponding Author) is with the School of Elec.  Eng., Tel-Aviv University, Tel-Aviv~69978, Israel. 
E-mail: \texttt{michaelm@tauex.tau.ac.il}}%
 }

\begin{document}

\maketitle

\begin{abstract}
We present a new sufficient condition for finite-gain $L_2$ input-to-output stability of a networked system.
The condition requires a matrix, that combines information  on the $L_2$ gains of the sub-systems and 
their interconnections, to be discrete-time diagonally stable~(DTDS). We show that the new result generalizes the standard  small gain theorem for the negative feedback connection of two sub-systems. An important advantage of the new result is that known sufficient conditions for DTDS
can be applied to derive sufficient conditions for networked input-to-output stability. We demonstrate this using several examples. We also derive a new necessary and sufficient condition for a matrix that is a rank one perturbation of a Schur diagonal matrix to be~DTDS. 
\end{abstract}

\begin{IEEEkeywords}
Input-to-output stability, networked systems, continuous-time diagonal stability. 
\end{IEEEkeywords}

\section{Introduction}

A powerful approach for
analyzing large-scale or networked systems is based on deducing properties of the networked system by  combining   properties of the sub-systems and their interconnection pattern. In particular, the input-output analysis   approach~\cite{vidya_large_scale_inter}  is based on combining input-output properties of the sub-systems and their interconnections to   
deduce input-output properties of the networked system. This approach  usually ignores the internal structure of the sub-systems that are described  as input-output operators, and thus 
yields   robustness to uncertainty in the dynamics and parameter values. 
Nevertheless,   under suitable  detectability and controllability conditions it is possible to deduce global asymptotic stability of the networked system (see,  e.g.,~\cite{SONTAG_secant_SCL}).

Important examples of the input-output
approach include:
(1)~the small gain theorem~\cite[Chapter~5]{khalil_book} that provides a sufficient condition for the   input-to-output stability of the negative feedback 
interconnection of two sub-systems;   (2)~a  condition that guarantees the passivity of a networked system based on
continuous-time 
diagonal stability of a matrix that combines 
information about the passivity properties of the sub-systems and their interconnection structure (see the elegant presentation in~\cite{ArcakSontag2008PassivityInterconnection}); 
and (3)~small gain conditions for  networked stability based 
on input-to-state stability, see e.g.~\cite{Jiang2018smallgainsurvey,liu2018nonlinear}.

Here, we present a  sufficient condition 
for~$L_2$ input-to-output stability of a   system composed of 
$n$ sub-systems interconnected via a linear network. 
 The condition requires a matrix, that combines the $L_2$ gains of the sub-systems and the interconnection strengths, to be  discrete-time diagonally
stable~(DTDS). We show that this condition is a generalization of the classical small-gain condition for~$L_2$ input-to-output  stability. 
We stress that while our condition is based on discrete-time diagonal stability, the result is applicable to both discrete-time and continuous-time networked systems.

The remainder of this note is organized as follows. The next section reviews known definitions and results that are used later on. 
Section~\ref{sec:main} presents the main result and its proof. An important advantage of the new result  is that it allows to use known conditions 
for~DTDS to derive conditions  for~$L_2$ input-to-output stability of a networked system. Section~\ref{sec:appli} demonstrates this using several applications. We also derive a new necessary and sufficient condition for a matrix that is a rank one perturbation of a Schur diagonal matrix to be~DTDS.  
The final section concludes and describes possible directions for further research.

We use standard notation. Small [capital] letters denote vectors [matrices]. For a vector~$v\in\R^n$, $\diag(v)$  is the~$n\times n$ diagonal matrix with~$v_i$ at entry~$(i,i)$. If~$v_i>0$ for all~$i$ 
then~$\diag(v)$ is called a positive diagonal matrix. 
The transpose of a matrix~$A$ is~$A^\top$, and~$|A|$ is the matrix obtained  from~$A$ be replacing every entry by its absolute value. For a square matrix~$A$, $\det(A)$ is the determinant of~$A$. 
The maximal  [minimal] eigenvalue of a symmetric matrix~$S$ is denoted by~$\lambda_{\max}(S)$ [$\lambda_{\min}(S)$].
A matrix~$P\in\R^{n\times n}$ is called positive-definite, denoted~$P\succ 0$,  if~$P$ is symmetric
and~$x^\top Px >0$ for all~$x\in\R^{n}\setminus\{0\}$. In this case,~$P^{1/2}$ is the unique positive-definite matrix such that~$P^{1/2}P^{1/2}=P$.
A matrix~$P\in\R^{n\times n}$ is called negative-definite, denoted~$P\prec 0$,  if~$-P$ is positive-definite.
We use $\|\cdot\| : \R^n \to \R_{\ge 0}$ to denote the Euclidean norm, and $\|A\| := \max_{\|x\|  = 1} \|Ax\|$ to denote the induced   matrix norm. Then~$\|A\|^2=\lambda_{\max}(A^\top A)$.
The non-negative orthant in~$\R^n$ is~$\R^n_{\geq 0}:=\{x \in\R^n \such x_i\geq 0 \text{ for all } i\}$.

The space of signals (thought of as time functions) $u : \R_{\ge0} \to \R^m$ which are piecewise continuous and satisfy 
\begin{align*}
\|  u\|_{T}& : =\left ( \int_0^T u^\top (t) u(t) \mathrm{d}t \right )^{1/2}\\
& <\infty  
\end{align*}
for all $T > 0$ is denoted $L_{2,e}^m$. 

\section{Preliminaries}
We begin by  quickly
reviewing several known definitions and results that will be used later on.
\subsection{Schur stability}
A matrix~$A\in\R^{n\times n}$ is called Schur   stable if~$|\lambda|<1$ for any eigenvalue~$\lambda$ of~$A$. 
This holds iff every solution of the discrete-time linear time-invariant~(LTI)
system
\be\label{eq:ds_lti}
x(k+1)=Ax(k)
\ee
converges to zero. 

A necessary and sufficient 
condition for Schur stability of~$A$ is that there exists a  solution~$P\succ 0 $ to the equation
\be\label{eq:atpap-p}
A^\top P A -P \prec 0. 
\ee
This implies that the function~$V:\R^n\to\R_{\geq 0 }$ defined by~$V(z):=z^\top P z$ is a Lyapunov  function for~\eqref{eq:ds_lti}. 

\begin{Remark}\label{rem:norm_scaled}
Note that multiplying~\eqref{eq:atpap-p} from the left  and
right by
$P^{-1/2}$  gives
\[
P^{-1/2}A^\top P  A
P^{-1/2} -  I_n\prec 0,
\]
that is,
\[
(P^{1/2} A P^{-1/2})^\top (P^{1/2} A P^{-1/2}) \prec I_n,
\]
so,
\[
\| P^{1/2} A P^{-1/2} \| <1.
\]
In other words, Schur stability is equivalent 
to a  weighted~$L_2$ norm of~$A$ being less than one.
\end{Remark}

    \begin{Remark}\label{rem:scale}
Let~$A\in\R^{n\times n}$ with~$A\not =0$. Fix a matrix~$P\succ 0$. 
Define
\[
\alpha:=\left(\frac{\lambda_{\min}(P)}{\lambda_{\max} (A^\top PA)}\right)^\frac1{2}.
\]
Then for any~$x\in\R^n\setminus\{0\}$  and~$c\in\R$ with~$|c|<\alpha$,
we have
\begin{align*}
    x^\top (cA^\top P c A)x  &\leq c^2 \lambda_{\max} (A^\top PA ) x^\top x\\
    &< \alpha^2 \lambda_{\max} (A^\top PA ) x^\top x\\
    &= \lambda_{\min}(P) x^\top x\\
    &\leq x^\top P x ,
\end{align*}
so~\eqref{eq:atpap-p} holds for the scaled matrix~$cA$.
\end{Remark}

\subsection{Discrete-time diagonal stability}
The matrix~$A \in\R^{n\times n} $ is called discrete-time diagonally stable~(DTDS) if there exists a positive diagonal matrix~$D$ such that
\be\label{eq:dad}
A^\top D A -D \prec 0,
\ee
or, equivalently, if there exists a diagonal Lyapunov function for~\eqref{eq:ds_lti}.

Remark~\ref{rem:scale} implies in particular that for any~$A\in\R^{n\times n}$ there exists~$\alpha>0$ such that the scaled matrix~$cA$ is~DTDS  for any~$|c|<\alpha$. In general,   easily verifiable necessary and sufficient conditions for~DTDS are not known. However,
  there exist several results for matrices    with a special structure. We list some of those  results   (see~\cite[Section 2.7]{diag_stab_book} for more details), as  combining them
  with the  new
   small  gain theorem in Section~\ref{sec:main} provides new
   explicit conditions for finite-gain $L_2$
   stability of a networked system.
\begin{Proposition}\label{prop:diag_stab_2x2}
    Let~$A \in \R^{2 \times 2}$. Then~$A$ is~DTDS iff   the following three
    inequalities hold:
    \begin{align*}
        |\det(A)| &< 1, \\
        |a_{11} + a_{22}| &< 1 + \det(A), \\
        |a_{11} - a_{22}| &< 1 - \det(A).
    \end{align*}
\end{Proposition}
For example, for~$A=\alpha I_2$  
these conditions become
$\alpha^2<1$, $|2\alpha|<1+\alpha^2$, and~$0<1-\alpha^2$, that is, $|\alpha|<1$. Indeed, for~$|\alpha|<1$, Eq.~\eqref{eq:dad} holds with~$D=I_2$, and for~$|\alpha|\geq 1$ we have that~$\alpha I_2$ is not Schur and thus not~DTDS. 

As another example, consider the case where~$A=uv^\top$, with~$u,v\in\R^2$. Then~$\det(A)=0$, so the conditions in Prop.~\ref{prop:diag_stab_2x2} become~$|u_1v_1+u_2v_2|<1$
and~$| u_1v_1-u_2v_2|<1 $. 
In this case, $0$ and~$u_1v_1+u_2v_2$ are the eigenvalues of~$A$, so the first condition  is a necessary and sufficient condition for Schur stability, and the additional condition is needed for the stronger property of~DTDS.


\begin{Proposition}\label{prop:DTDS_SYMMETRIC}
    Let~$A \in \R^{n \times n}$. If~$A$ is Schur 
    and there exists a non-singular diagonal matrix~$\Omega\in\R^{n\times n}$ such that~$\Omega A \Omega^{-1} $ is symmetric then $A$ is~DTDS.
\end{Proposition}
In particular, if~$A$ is Schur and symmetric then it is~DTDS.

\begin{Proposition}\label{prop:DTDS_similarity}
    Let~$A \in \R^{n \times n}$. If~$D$ is a positive diagonal matrix then~$DAD^{-1}$ is~ DTDS iff~$A$ is~DTDS.
\end{Proposition}

\begin{Proposition}\label{prop:DTDS_nonneg}
    Let~$A \in \R^{n \times n}$. If~$|A|$ is Schur then $A$ is DTDS.
\end{Proposition}
\begin{Example}\label{exa:22zerodiag}
    Suppose that~$A\in\R^{2\times 2}$ with~$a_{11}=a_{22}=0$. 
In this case, it is straightforward to verify that~$A$ is Schur iff~$|A|$ is Schur, so
\[
A \text{ is Schur } \iff |A| \text{ is Schur } \iff A 
\text{ is DTDS}.
\]
\end{Example}

\begin{Remark}\label{rem:dts+zeros}
    Prop.~\ref{prop:DTDS_nonneg} implies in particular that when all entries of~$A$ are non-negative, then~$A$ is Schur iff it is DTDS (see also~\cite{Pastravanu2006GenDiagStab}). However, in general, 
\be\label{eq:strict}
 \{A \such |A| \text{ is Schur}\} \text{ is strictly contained in } \{A \such 
A \text{ is DTDS}\}.
\ee
For example, the matrix
\begin{equation*}
    A = \frac1{2}\begin{bmatrix}
        -1 & 1 \\
        -1 & -1
    \end{bmatrix}.
\end{equation*}
 is DTDS by Prop.~\ref{prop:diag_stab_2x2}, 
 but~$|A|$ is not Schur.  
As another example, the matrix
\begin{equation*}
    A = \begin{bmatrix}
        0 & 0.23 & 0.56 & 0.56 \\
        0.51 & 0 & 0.56 & 0.09 \\
        -0.27 & -0.12 & 0 & 0.4 \\
        0.51 & 0.15 & 0.57 & 0
    \end{bmatrix}
\end{equation*}
is DTDS (with~$D = \diag(0.9994, 0.585, 1.8213, 0.9629)$), but~$|A|$ is not Schur.
\end{Remark}

The property described in~\eqref{eq:strict} implies  that, unlike the standard small-gain theorem, our main
result (Theorem~\ref{thm:smallgain_gen} below) takes into account both the gain
and the phase  of the interconnections in the networked system.

\section{A condition for input-to-output stability of a  networked system}\label{sec:main} 

Consider a networked
system consisting of $n$ sub-systems considered as operators~$G_i : L_{2,e}^{m_i} \to L_{2,e}^{m_i}, i = 1,\dots,n$,
with input $u_i : [0,\infty) \to \R^{m_i}$ and output $y_i : [0,\infty) \to \R^{m_i}$. 
We assume that all
the sub-systems are  finite-gain  $L_2$  stable, i.e., there exist~$\gamma_i > 0$ and~$\beta_i \ge 0$ such that for any input~$u_i\in L_{2,e}^{m_i}$ and corresponding output~$y_i$, we have 
\be\label{eq:yi_ui}
        \|  {y_i}  \|_{T} \leq  \gamma_i \|{u_i}\|_{T}
        + \beta_i \text{ for all } T > 0.
\ee
Define~$m := \sum_{i=1}^n m_i$, and 
let
\[
y:= \begin{bmatrix} y_1^\top & \dots & y_n^\top \end{bmatrix}^\top.
\]
The sub-systems are connected to each other in a linear form 
via
\be\label{eq:uilin}
u_i = v_i + \sum_{j=1}^n A_{ij} y_j, 
\ee
where  $v_i : \R_{\ge0} \to \R^{m_i}$ is an external input,  and $A_{ij} \in \R^{m_i \times m_i}$, $i=1,\dots,n$. 
We assume that the resulting networked system is well-posed in the sense of~\cite[Chapter~2]{vidya_large_scale_inter}.
In particular, for any set of inputs~$v_i \in L_{2,e}^{m_i}$, $i=1,\dots,n$,
there exists a unique output~$y\in L_{2,e}^{m}$.

Define  the interconnection  matrix~$A\in\R^{m \times m}$ by 
\begin{equation}
    A := \begin{bmatrix}
        A_{11} & \cdots & A_{1n} \\
        \vdots & \ddots & \vdots \\
        A_{n1} & \cdots & A_{nn}
    \end{bmatrix}.
\end{equation}

We can now state our main result that provides a sufficient condition for   finite-gain  $L_2$ stability of the networked system.
\begin{Theorem}\label{thm:smallgain_gen}
    Let $\Gamma := \diag(\gamma_1 I_{m_1}, \dots, \gamma_n I_{m_n})$. Suppose that there exist~$d_1,\dots,d_n>0$ such  that
    \be\label{eq:kron}
    A^\top \Gamma D \Gamma A \prec D
    \ee
    where~$D := \diag(d_1 I_{m_1}, \dots, d_n I_{m_n})$.
     Then there exist  $\rho,\beta  > 0$  such that 
  \[
            \| y\|_T\leq \rho \|  v \|_T + \beta,
  \]
    for all~$v \in L_{2,e}^{m}$ and all~$T>0$.
\end{Theorem}

\begin{Remark}
    In the single-input single-output case, i.e. when~$m_1=\dots=m_n=1$, we have~$D=\diag(d_1,\dots,d_n)$, and   condition~\eqref{eq:kron}  reduces to~$\Gamma A$ being~DTDS. If~$m_i>1$ for some $i \in \{1,\dots,n\}$ then~$D=\diag(d_1 I_{m_1},\dots,d_n I_{m_n})$ is still a diagonal matrix, but it has a special block-diagonal
    structure, namely,
    it has~$n$ diagonal blocks, and  
    block~$i$ has the form~$d_i I_{m_i}$.
    Thus, when~$m_i>1$  for some~$i$ condition~\eqref{eq:kron}  requires~$\Gamma A$ to   be ``block~DTDS''. 
\end{Remark}

\begin{Remark}
Classic generalizations of the small-gain theorem are based on algebraic properties of a ``test matrix''~\cite[Chapter~6]{vidya_large_scale_inter}. However, every entry in such a matrix is the \emph{gain}
of a sub-system and/or interconnection operator and in particular each entry is non-negative. Thus, the stability condition ignores the ``phase'' of the interconnections.
Theorem~\ref{thm:smallgain_gen} uses the matrix~$A$ that may include negative entries, and thus takes into account both the magnitude and the sign of the entries.
\end{Remark}

\begin{IEEEproof}[Proof of Theorem~\ref{thm:smallgain_gen}]
Assume that~\eqref{eq:kron} holds. Let~$D^{1/2}$ denote the positive square root of~$D$.
        Since $A^\top \Gamma D \Gamma A - D \prec 0$, Remark~\ref{rem:norm_scaled} implies  that $1-\|D^{1/2}\Gamma A D^{-1/2}\| > 0$. Fix~$\varepsilon>0$ sufficiently small such that
        \[
        s:=1-\|D^{1/2}\tilde \Gamma A D^{-1/2}\| > 0,
        \]
        where~$\tilde \Gamma:=\diag(\tilde \gamma_1 I_{m_1},\dots,
        \tilde \gamma_n I_{m_n}  )$, with~$\tilde \gamma_i :=\gamma_i \sqrt{1+\varepsilon}  $    .

Eq.~\eqref{eq:yi_ui} gives
\begin{align*}
    \|  {y_i}  \|_{T}^2& \leq  \gamma_i ^2 \|{u_i}\|_{T}^2+2\gamma_i\beta_i \|u_i\|_T
        + \beta_i^2\\
&\leq \tilde\gamma_i ^2\|{u_i}\|_{T}^2+q_i^2,
\end{align*}
with~$q_i:=\sqrt{\frac{(1+\varepsilon)\beta_i^2+\varepsilon^2}{\varepsilon} }$, for all~$T\geq 0$.
Now,
 \begin{align*}
  \|D^{1/2} y  \|_T^2 & =
  \int_0^T  \sum_{i=1}^n d_iy_i^\top (t) y_i(t) dt 
  \\
  &\leq  
  \sum_{i=1}^n 
(d_i \tilde\gamma_i^2\|u_i\|_T^2 +d_iq_i ^2 )
 \\&=
  \|D^{1/2} \tilde\Gamma  u   \|_T^2+r^2, 
\end{align*}
where~$r:=\sqrt{\sum_{i=1}^n d_iq_i^2}$. Thus, 
 \begin{align*}
  \|D^{1/2} y  \|_T 
  &\leq
  \|D^{1/2} \tilde\Gamma u   \|_T+r \\
  &=\| D^{1/2} \tilde\Gamma v +D^{1/2}\tilde\Gamma A y        \|_T+r
  \\ &\leq 
  \| D^{1/2} \tilde\Gamma v      \|_T
  +\|  D^{1/2}  \tilde\Gamma  A D^{-1/2}D^{1/2} y        \|_T+r\\
  &\leq  \| D^{1/2} \tilde\Gamma v      \|_T
  +\|D^{1/2}  \tilde\Gamma  A D^{-1/2} \| \|  D^{1/2} y        \|_T +r. 
\end{align*}
   Rearranging this and using the fact that diagonal matrices commute gives
\begin{align*}
s  \|  D^{1/2} y        \|_T \leq 
 \|\tilde\Gamma\| \|D^{1/2} v\|_T +r. 
\end{align*}
 Let~$d_m,d_M>0$ denote the minimal and maximal diagonal entries of~$D$. Then
 \[
s d_m^{1/2} \|y\|_T \leq   \|\tilde\Gamma\| d_M^{1/2}\|v\|_T +r,  
 \]   
for any~$T\geq 0$,  
and
this    completes the proof. 
\end{IEEEproof}

\begin{Remark}
Theorem~\ref{thm:smallgain_gen} can be easily applied to a networked system composed of discrete-time sub-systems. In this case, the inputs and outputs are discrete-time signals, i.e., $y_i,v_i : \N \to \R^{m_i}$ and the integrals in the norms of the signals are replaced by sums. Similarly, Theorem~\ref{thm:smallgain_gen} can also   be restated in terms of \emph{incremental} finite gain stability~\cite{Zames1966}, that is, assuming that the sub-systems satisfy
\begin{equation*}
    \|y_i - \tilde y_i\|_T \le \gamma_i \|u_i - \tilde u_i\|_T
\end{equation*}
for all pairs of inputs~$u_i,\tilde u_i$ and corresponding outputs~$y_i,\tilde y_i$, the same DTDS condition implies
that the networked system is incrementally stable.
\end{Remark}

\begin{figure}
    \centering
    \begin{subfigure}{0.45\textwidth}
        \begin{tikzpicture}[
            block/.style = {draw, rectangle, thick, minimum height=2em, minimum width=3em},
            sum/.style = {draw, circle, thick, minimum width=1.5em}]
            
            \node[block, minimum height=1cm] (sys1) {$G_1$} ;
            \node[block, minimum height=1cm] (sys2) [below=0.3cm of sys1] {$G_2$} ;
            \node[sum, left=0.65cm of sys1] (fbsum1) {};
            \node[sum, right=0.65cm of sys2] (fbsum2) {};
    
            \coordinate (sys1aux) at (sys1 -| fbsum2);
            \coordinate (sys2aux) at (sys2 -| fbsum1);
            
            \draw[->, thick] (sys1.east) -| (fbsum2.north);
            \draw[->, thick] (sys1aux) -- +(1,0) node[right] {$y_1$};
    
            \draw[->, thick] (fbsum2.west) -- (sys2.east);
            \draw[->, thick] ({$(sys1aux)+(1,0)$} |- fbsum2) node[right] {$v_2$} -- (fbsum2.east);
    
            \draw[->, thick] (sys2.west) -| (fbsum1.south);
            \draw[->, thick] (sys2aux) -- +(-1,0) node[left] {$y_2$};
    
            \draw[->, thick] (fbsum1.east) -- (sys1.west);
            \draw[->, thick] ({$(sys2aux)+(-1,0)$} |- fbsum1) node[left] {$v_1$} -- (fbsum1.west);
    
            \draw[->, thick] (sys1aux) -- +(0,1) -| (fbsum1.north);
            \draw[->, thick] (sys2aux) -- +(0,-1) -| (fbsum2.south);

            \node[] () at ($(fbsum1.south) + (0,0.1)$){${\scriptscriptstyle -}$};
            \node[] () at ($(fbsum1.west) + (0.12,0)$){${\scriptscriptstyle +}$};
            \node[] () at ($(fbsum1.north) - (0,0.1)$){${\scriptscriptstyle -}$};

            \node[] () at ($(fbsum2.north) - (0,0.12)$){${\scriptscriptstyle +}$};
            \node[] () at ($(fbsum2.east) - (0.14,0)$){${\scriptscriptstyle +}$};
            \node[] () at ($(fbsum2.south) + (0,0.1)$){${\scriptscriptstyle -}$};
        \end{tikzpicture}
        \label{fig:fb}
    \end{subfigure}
    \begin{subfigure}{0.45\textwidth}
        \begin{tikzpicture}[
            block/.style = {draw, rectangle, thick, minimum height=2em, minimum width=3em},
            sum/.style = {draw, circle, thick, minimum width=1.5em}]
            
            \node[block, minimum height=1cm] (sys1) {$\frac{\gamma_1}{1 - \gamma_1}$} ;
            \node[block, minimum height=1cm] (sys2) [below=0.3cm of sys1] {$\frac{\gamma_2}{1 - \gamma_2}$} ;
            \node[sum, left=0.65cm of sys1] (fbsum1) {};
            \node[sum, right=0.65cm of sys2] (fbsum2) {};
    
            \coordinate (sys1aux) at (sys1 -| fbsum2);
            \coordinate (sys2aux) at (sys2 -| fbsum1);
            
            \draw[->, thick] (sys1.east) -| (fbsum2.north);
            \draw[->, thick] (sys1aux) -- +(1,0) node[right] {$y_1$};
    
            \draw[->, thick] (fbsum2.west) -- (sys2.east);
            \draw[->, thick] ({$(sys1aux)+(1,0)$} |- fbsum2) node[right] {$v_2$} -- (fbsum2.east);
    
            \draw[->, thick] (sys2.west) -| (fbsum1.south);
            \draw[->, thick] (sys2aux) -- +(-1,0) node[left] {$y_2$};
    
            \draw[->, thick] (fbsum1.east) -- (sys1.west);
            \draw[->, thick] ({$(sys2aux)+(-1,0)$} |- fbsum1) node[left] {$v_1$} -- (fbsum1.west);

            \draw[->, thick, draw opacity=0] (sys1aux) -- +(0,1) -| (fbsum1.north);
            \draw[->, thick, draw opacity=0] (sys2aux) -- +(0,-1) -| (fbsum2.south);

            \node[] () at ($(fbsum1.south) + (0,0.1)$){${\scriptscriptstyle -}$};
            \node[] () at ($(fbsum1.west) + (0.12,0)$){${\scriptscriptstyle +}$};

            \node[] () at ($(fbsum2.north) - (0,0.12)$){${\scriptscriptstyle +}$};
            \node[] () at ($(fbsum2.east) - (0.14,0)$){${\scriptscriptstyle +}$};
        \end{tikzpicture}
        \label{fig:fb_sg}
    \end{subfigure}
    \caption{Block diagram of a feedback interconnection of two sub-systems (top),  and a simplified diagram (bottom). \label{fig:leftright}}
\end{figure}
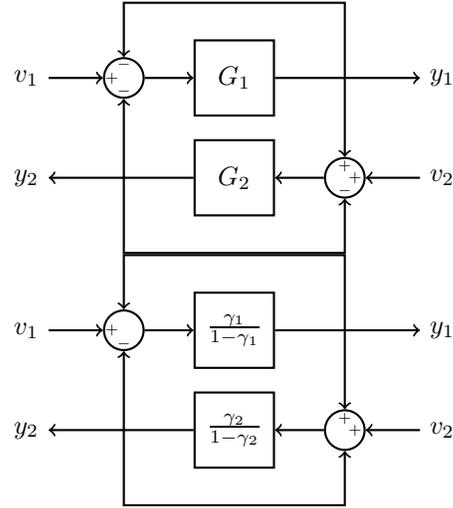

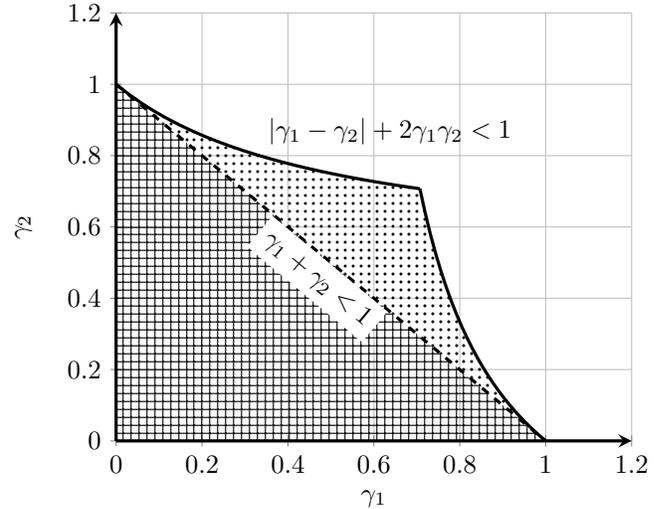
\begin{figure}
    \centering
    \begin{tikzpicture}
        \begin{axis}[%
            axis line style=very thick, 
            axis x line=bottom, 
            axis y line=left, 
            ymin=0,ymax=1.2,xmin=0,xmax=1.2, 
            xlabel=$\gamma_1$, ylabel=$\gamma_2$,grid=major 
        ]
        \addplot[name path=stdsg, very thick, domain=0:1, dashed] {1-x} node[pos=.5,below,sloped,fill=white] {$\gamma_1 + \gamma_2 < 1$}; 
        \addplot[name path=oursg1, very thick, domain=0:0.707] {(1+x)/(1+2*x)} node[pos=.5,above right] {$|\gamma_1 - \gamma_2| + 2 \gamma_1 \gamma_2 < 1$}; 
        \addplot[name path=oursg2, very thick, domain=0.707:1] {(x-1)/(1-2*x)}; 

        \path[name path=axis] (axis cs:0,0) -- (axis cs:1,0);

        \addplot [
            thick,
            color=blue,
            fill=blue, 
            fill opacity=0.75,
            pattern=grid
        ]
        fill between[
            of=stdsg and axis,
            soft clip={domain=0:1},
        ];

        \addplot [
            thick,
            color=blue,
            fill=blue, 
            fill opacity=0.75,
            pattern=dots
        ]
        fill between[
            of=oursg1 and stdsg,
            soft clip={domain=0:0.707},
        ];
        \addplot [
            thick,
            color=blue,
            fill=blue, 
            fill opacity=0.75,
            pattern=dots
        ]
        fill between[
            of=oursg2 and stdsg,
            soft clip={domain=0.707:1},
        ];
      \end{axis} 
    \end{tikzpicture}
    \caption{Comparing two sufficient conditions for netwroked 
    stability derived in Example~\ref{exa:gene22}.}
    \label{fig:exa_stab_region}
\end{figure}

\begin{Example}\label{exa:gene22}
    Suppose that~$n=2$ and~$m_1=m_2=1$, i.e. 
  the networked  system consists of two single-input single-output sub-systems, and  assume  a general
  feedback configuration  
\be 
    A = \begin{bmatrix}
        a_{11} & a_{12} \\
        a_{21} & a_{22}
    \end{bmatrix}. 
\ee 
Taking~$\Gamma=\diag(\gamma_1,\gamma_2)$, with~$\gamma_i>0$, we have
\begin{equation*}
    \Gamma A =  \begin{bmatrix}
        \gamma_1 a_{11} & \gamma_1 a_{12} \\
        \gamma_2 a_{21} & \gamma_2 a_{22}
    \end{bmatrix}.
\end{equation*}
By Prop.~\ref{prop:diag_stab_2x2}, $\Gamma A$ is~DTDS  
iff   the following three
    inequalities hold:
    \begin{align}\label{eq:3xcon}
        \gamma_1\gamma_2 |\det(A)| &< 1, \nonumber\\
        |\gamma_1 a_{11} + \gamma_2 a_{22}| &< 1 +\gamma_1\gamma_2  \det(A), \\
        |\gamma_1 a_{11} -\gamma_2  a_{22}| &< 1 - \gamma_1\gamma_2 \det(A).\nonumber
    \end{align}
We conclude that if these three conditions hold then the networked system
is finite-gain $L_2$ stable. 
We consider two more concrete examples.

First, assume that~$a_{11}=a_{22}=0$, and~$a_{21}=1$, so
\be 
    A = \begin{bmatrix}
        0 & a_{12} \\
        1 & 0
    \end{bmatrix},
\ee
Then~$\Gamma A$ is DTDS iff~$|\det(\Gamma A)| = |a_{12}| \gamma_1 \gamma_2 < 1$. If~$a_{12}\in\{-1,1\}$ then the condition  becomes~$\gamma_1\gamma_2<1$, so in particular this recovers the standard small-gain theorem for the negative feedback interconnection of two finite-gain sub-systems~\cite[Chapter 5]{khalil_book}. The reason that in this case the condition based on~DTDS is as conservative
as the small-gain theorem is explained in Example~\ref{exa:22zerodiag}.

Second, suppose that
$a_{11}=a_{12}=a_{22}=-1$, and~$ a_{21}=1$, that is,
\be\label{eq:afedd}
A=\begin{bmatrix}
    -1&-1\\1&-1
\end{bmatrix}.
\ee
Then the three conditions in~\eqref{eq:3xcon} become 
\begin{align*}
        \gamma_1 \gamma_2 &< 1/2,    \\
        \gamma_1 + \gamma_2 &< 1 + 2\gamma_1 \gamma_2,    \\
        |\gamma_1 - \gamma_2| &< 1 - 2\gamma_1 \gamma_2. 
    \end{align*}
    It is clear that the third condition implies the first one, and some algebra shows that the third condition also implies the second one. We conclude that  the networked
    system is finite-gain $L_2$ stable if
    \begin{equation}\label{eq:ga1ga2}
        |\gamma_1 - \gamma_2| + 2\gamma_1 \gamma_2 < 1.
    \end{equation}
    To compare this with the bound  derived  using the small gain theorem, note that we can depict the system with interconnection matrix in~\eqref{eq:afedd} as in Fig.~\ref{fig:leftright} (top), where~$G_i$ has a gain~$\gamma_i$, and this can be converted to the closed-loop system depicted in Fig.~\ref{fig:leftright} (bottom).
Applying the small gain theorem to the closed-loop feedback system implies that a sufficient condition for finite gain  stability
is~$\frac{\gamma_1}{1-\gamma_1} \frac{\gamma_2}{1-\gamma_2}<1 $, that is,~$\gamma_1+\gamma_2<1$.  This condition is more conservative than~\eqref{eq:ga1ga2}, see Fig.~\ref{fig:exa_stab_region}.
\end{Example}

\subsection{The case of rank one connections}
Consider the case where all the sub-systems are SISO (that is,~$m_i=1$ for all~$i$) and
are  interconnected via  
\be\label{eq:rank1_conn}
u_i = v_i + s_i y_i+  k_i g^\top y,
\ee
with~$g\in\R^n$ and~$k_i,s_i\in\R$.  Here,~$g^\top y$ may be interpreted as a ''weighted average'' of all the outputs, which is fed to all the sub-systems, albeit with   different gains~$k_i$. In many cases,~$g\in\R^n_{\geq 0}$. As noted in~\cite{SimpsonPorco2022DiagStabAGC}, this form of interconnection appears in many applications including: optimal frequency control~\cite{DORFLER_gather_and_broadcast},
automatic generation control in power systems~\cite{auto_power_gener}, 
game-theoretic CDMA power control~\cite{Passive_CDMA},
and consensus  dynamics~\cite{multiagent_netwroks_book}.

The connection matrix~$A\in\R^{ n\times n  }$ corresponding to~\eqref{eq:rank1_conn}
is~$    A = \diag(s_1,\dots,s_n)+ kg^\top
   $, with~$k:=\begin{bmatrix}
        k_1&\dots&k_n
    \end{bmatrix}^\top$,
and thus
\be\label{eq:GA_rank1}
\Gamma A=  \diag(\gamma_1s_1,\dots,\gamma_n s_n)+\Gamma k g^\top 
\ee
is a rank one perturbation of a diagonal matrix. The next result allows using Theorem~\ref{thm:smallgain_gen} to analyze this case, by providing  a necessary and sufficient condition for a matrix that is the sum of a Schur diagonal matrix and a rank one matrix to be~DTDS.
To the best of our knowledge, this result  is novel and may be of independent interest. 

\begin{Theorem}\label{thm:DS_rank_1}
Consider the matrix
\begin{equation}
    A = \Delta + u v^\top,
\end{equation}
with~$ \Delta = \diag(\delta_1, \dots, \delta_n)\in\R^{n\times n}$,  where~$|\delta_i| < 1, i=1,\dots,n$,   $u \in \R^n $, and~$v \in \R_{\ge0}^n$.
Suppose that~$A$ is Schur. Then~$A$ is~DTDS iff 
\be\label{eq:sum_cond}
\sum_{i=1}^n \frac1{1-\delta_i^2} [  cu_i v_i  ]_+ < 1,  
\ee
where~$[z]_+:=\max\{z,0\}$, and
\be\label{+eq:def_c_rank1}
c:=\frac{-2}{1+ v^\top (\Delta-I_n)^{-1} u}.
\ee.
\end{Theorem}
The proof is placed in the Appendix. 

\begin{Corollary}\label{coro:rank_one_DTDS}
    Consider the rank one matrix
    \begin{equation}
        A = uv^\top
    \end{equation}
    where~$u,v \in \R^n$.   Then~$A$
    is~DTDS 
      iff~$|v|^\top |u|<1$ .
\end{Corollary}
\begin{IEEEproof}
We first show that~$uv^\top$ is~DTDS iff~$|u|v^\top $ is~DTDS iff~$|u||v^\top|$ is~DTDS. 
To show this, note that~$uv^\top $ is~DTDS iff there exists a positive diagonal matrix~$D$ such that
\begin{align*}
    D& \succ (uv^\top)^\top D (uv^\top)\\
    &=(u^\top D u)v v^\top.
\end{align*}
Clearly, the last term does not change if we replace~$u_i$ by~$|u_i|$ for any~$i$. Since~$A$ is DTDS iff~$A^\top$ is DTDS, a similar argument shows that~DTDS of~$uv^\top$ is equivalent to~DTDS of~$u
|v^\top|$. 
Thus, we may assume in the remainder  of the proof that
$u,v\in\R^n_{\geq 0}$. 
Since~$v^\top u$ is an eigenvalue of~$A$,~$v^\top u <1$ is a necessary condition for~DTDS.  
   Now applying Theorem~\ref{thm:DS_rank_1} with~$\Delta = 0$, we have that~$A$ is DTDS iff
    \begin{align*}
        1 &> \sum_{i=1}^n [\frac{-2 u_i v_i}{1 - v^\top u}]_+ \\
          &= \frac{2}{1 - v^\top u} \sum_{i=1}^n [-v_i  u_i ]_+\\&=0,
    \end{align*}
    where we used the fact that~$v_i u_i \geq 0$ for all~$i$, and that~$1-v^\top u >0$.
    Thus, when~$v,u\in\R^n_{\geq 0} $
the matrix~$A$ is~DTDS iff~$v^\top u<1$,
    and this completes the proof.
\end{IEEEproof}

\section{Applications}\label{sec:appli}
An important advantage  of Theorem~\ref{thm:smallgain_gen}
is that 
known results on DTDS can now  be immediately 
used to derive  conditions  guaranteeing finite-gain $L_2$ stability of  the  networked  system. The next      result  demonstrates this. 

\begin{Corollary}
Consider the networked system with SISO sub-systems.
If any one of the following conditions holds then the networked system is finite-gain $L_2$ stable. 
\begin{enumerate}
    \item The gains satisfy $\gamma_i\leq 1$ for all~$i$ and~$A$ is~DTDS;
    \item All the entries of~$A$ are non-negative and~$\Gamma A$ is Schur; 
    \item The sub-systems are identical, $A$ is symmetric and~$\Gamma A$ is Schur;
    \item $A$ is triangular and $\Gamma A$ is Schur;
    \item The interconnection pattern is as in~\eqref{eq:rank1_conn}, with~$|\gamma_i s_i|<1$ for all~$i$, and
    Eqs.~\eqref{eq:sum_cond} and
    \eqref{+eq:def_c_rank1}
 hold with~$\delta_i=\gamma_i s_i$, $u=\Gamma k $, and~$v=g\in\R^n_{\geq 0}$.
\end{enumerate}
\end{Corollary}
 
\begin{IEEEproof}
Recall that if~$A$ is DTDS and~$\Gamma$ is  a  positive diagonal matrix with all entries smaller or equal to
  one then~$\Gamma A$ is also DTDS~\cite{BHAYA1993}. Combining this  with Theorem~\ref{thm:smallgain_gen} proves 1). The proof of 2) follows from combining Prop.~\ref{prop:DTDS_nonneg}   and  Theorem~\ref{thm:smallgain_gen}.  
To prove 3), note that if $A$ is symmetric and all the sub-systems are identical then~$\Gamma A=\gamma I_n A$ is symmetric and if it is also Schur then Prop.~\ref{prop:DTDS_SYMMETRIC} implies that it is~DTDS. The proof of 4) follows  from the fact that a triangular Schur matrix is~DTDS~\cite{BHAYA1993},
and the proof of~5) follows  from Thm.~\ref{thm:DS_rank_1}. 
\end{IEEEproof}
Note that some of these results have a clear control-theoretic interpretation in terms of the networked system, for example, statement 4) corresponds to a serial interconnection of the sub-systems in the network. 

\section{Discussion}
We derived a new condition guaranteeing   finite-gain   $L_2$ stability of a networked system. 
This is based on the DTDS of the matrix~$\Gamma A$, where~$\Gamma$ is a diagonal matrix collecting the gains of the sub-systems, and~$A$ is a matrix describing the interconnections
of the sub-systems that may have arbitrary signs. 
We showed  that the standard small gain theorem is a special case of the new condition and, furthermore, that known results on DTDS can be used to derive sufficient conditions for   finite-gain stability of a networked system.  In particular, we derived a new necessary and sufficient condition for DTDS of matrix that is a  rank one perturbation of a   diagonal Schur matrix, and applied it to analyze   finite-gain  $L_2$ stability of a networked system with a specific structure. 

We believe that Theorem~\ref{thm:smallgain_gen} suggests many interesting research directions. First, there are many more sufficient conditions for~DTDS 
(see, e.g.~\cite{BHAYA1993}), 
and it may be interesting  to study their implications in the context of Theorem~\ref{thm:smallgain_gen}. Second, there exist conditions guaranteeing  that a matrix remains~DTDS after a perturbation (see,  e.g., \cite[Chapter~2]{diag_stab_book}) and it may  be interesting  to interpret such results in the framework of robustness of  networked systems. Finally,  small gain results can also be used for control synthesis.

\section*{Appendix: Proof of Theorem~\ref{thm:DS_rank_1}}
The proof is based on converting the problem of determining~DTDS to the problem of determining continuous-time diagonal stability~(CTDS), and then applying the results in~\cite{SimpsonPorco2022DiagStabAGC} on~CTDS of a matrix that is a rank-one perturbation of a negative diagonal matrix.

Given~$A\in\R^{n\times n}$ satisfying that~$1$ is not an eigenvalue of~$A$, define~$\tilde A\in\R^{n\times n}$ via the bilinear transformation:
\[
 \tilde A := (A + I_n)(A - I_n)^{-1} .
\]
Suppose that~$P\in\R^{n\times n}$ is symmetric. Then
\begin{align*}
    P\tilde A+\tilde A^\top P& = P(A + I_n)(A - I_n)^{-1}\\&+
    (A^\top - I_n)^{-1}
    (A ^\top+ I_n)  P,
\end{align*}
so
\begin{align*}
(A-I_n)^\top  (   P\tilde A+\tilde A^\top P) (A-I_n)& = (A^\top-I_n) P(A + I_n)\\&+
    (A ^\top+ I_n)  P (A-I_n)\\&=2(A^\top P A-P).
\end{align*}
In particular, $P\tilde A+\tilde A^\top P \prec 0 $  iff
$A^\top PA-P\prec 0$.
Also, this implies that if~$1$ is not an eigenvalue of~$A$ then~$A$ is DTDS iff~$\tilde A$ is continuous-time diagonally stable~(CTDS).

Now consider the matrix
\begin{equation}
    A = \Delta + u v^\top,
\end{equation}
with~$ \Delta = \diag(\delta_1, \dots, \delta_n)\in\R^{n\times n}$,  where~$|\delta_i| < 1$, $i=1,\dots,n$, and $u \in \R^n ,v \in \R_{\ge0}^n$. Then
\begin{align*}
    \det(I_n - A) &= \det (   (I_n-\Delta)(I_n-(I_n-\Delta)^{-1} uv^\top    )\\  
         &= \det(I_n - \Delta) (1 - v^\top (I_n - \Delta)^{-1} u) . 
\end{align*}
In particular, if~$v^\top (I_n-\Delta)^{-1} u = 1$
then~$1$ is an eigenvalue of~$A$, so~$A$ is not Schur 
and thus not~DTDS.  Thus, we assume from here on that
\[
v^\top (I_n-\Delta)^{-1} u \not = 1. 
\]
 Now, 
\begin{align*}
    \tilde A &= (A + I_n)(A - I_n)^{-1} \\
      &= (2I_n+\Delta - I_n + u v^\top)(\Delta - I_n + u v^\top)^{-1} \\
      &=I_n+ 2(\Delta - I_n + u v^\top)^{-1} .
\end{align*}
By the
Sherman–Morrison formula, 
\begin{align}\label{eq:go_bakla}
    \tilde A &= 
   I_n+2(\Delta-I_n)^{-1}
    -2\frac{(\Delta-I_n)^{-1}  u v^\top (\Delta-I_n)^{-1}}{1+ v^\top (\Delta-I_n)^{-1} u}.
\end{align}
Thus, 
\begin{align*}
    (\Delta-I_n) \tilde A (\Delta-I_n)=-S -\frac{2uv^\top}{1+ v^\top (\Delta-I_n)^{-1} u}.
\end{align*}
with~$S:=\diag(1-\delta_1^2,\dots, 1-\delta_n^2)$.
Since the property of~CTDS is invariant under pre- and post-multiplication by positive diagonal matrices, we have that~$\tilde A$ is~CTDS iff
$-S + c uv^\top$ is~CTDS, where
\[
c:=\frac{-2}{1+ v^\top (\Delta-I_n)^{-1} u}.
\]
 Using~\cite[Theorem II.1]{SimpsonPorco2022DiagStabAGC}  we   get that~$\tilde A$ is~CTDS if and only if
\[
\sum_{i=1}^n \frac1{1-\delta_i^2} [  cu_i v_i  ]_+ < 1,  
\]
where~$[z]_+:=\max\{z,0\}$, and this completes the proof.~\hfill{\qed}

\bibliographystyle{IEEEtranS}
\bibliography{fullrefs}

\end{document}